\documentclass[aps, prl, reprint, showpacs, superscriptaddress]{revtex4-1}

\usepackage{graphicx}
\usepackage{dcolumn}
\usepackage{amsmath}
\usepackage{subfigure}
\usepackage{color}
\usepackage{float}

\begin{document}

\title{Magnetically Charged Superdomain Walls In Square Artificial Spin Ice}

\author{J. C. T Lee}
\thanks{These authors contributed equally to this work.}
\affiliation{Advanced Light Source, Lawrence Berkeley National Laboratory, Berkeley, CA 94720-8229, USA}
\affiliation{Department of Physics, University of Oregon, Eugene, OR 97403-1274, USA}

\author{S. K. Mishra}
\thanks{These authors contributed equally to this work.}
\affiliation{Advanced Light Source, Lawrence Berkeley National Laboratory, Berkeley, CA 94720-8229, USA}

\author{V. S. Bhat} 
\affiliation{Department of Physics and Center for Advanced Materials, University of Kentucky, Lexington, KY 40506-0055, USA}

\author{R. Streubel} 
\affiliation{Materials Sciences Division, Lawrence Berkeley National Laboratory, Berkeley, CA 94720-8229, USA}

\author{B. Farmer} 
\affiliation{Department of Physics and Center for Advanced Materials, University of Kentucky, Lexington, KY 40506-0055, USA}

\author{X. Shi}
\affiliation{Advanced Light Source, Lawrence Berkeley National Laboratory, Berkeley, CA 94720-8229, USA}
\affiliation{Department of Physics, University of Oregon, Eugene, OR 97403-1274, USA}

\author{L. E. De Long}
\affiliation{Department of Physics and Center for Advanced Materials, University of Kentucky, Lexington, KY 40506-0055, USA}

\author{I. McNulty}
\affiliation{Center for Nanoscale Materials, Argonne National Laboratory, Argonne, IL 60439, USA}

\author{P. Fischer} 
\affiliation{Materials Sciences Division, Lawrence Berkeley National Laboratory, Berkeley, CA 94720-8229, USA}
\affiliation{Department of Physics, University of California, Santa Cruz, CA 95064, USA}

\author{S. D. Kevan} 
\affiliation{Advanced Light Source, Lawrence Berkeley National Laboratory, Berkeley, CA 94720-8229, USA}
\affiliation{Department of Physics, University of Oregon, Eugene, OR 97403-1274, USA}

\author{S. Roy}
\email[Corresponding author: ]{sroy@lbl.gov}
\affiliation{Advanced Light Source, Lawrence Berkeley National Laboratory, Berkeley, CA 94720-8229, USA}

\date{\today}
 
\begin{abstract}

We report direct evidence that magnetically charged superdomain walls form spontaneously in two dimensional square artificial spin ice nanostructures in response to external magnetic fields.
These extended magnetic defects were revealed by the development of internal structure, which varies as a function of applied magnetic field, within the Bragg peaks of resonant soft x-ray magnetic scattering patterns.
Magnetically charged superdomain walls extend over tens of lattice sites and do not necessarily align with the applied field.  
Our results illustrate a novel approach to detect hierarchical magnetic structures within spin textures.

\end{abstract}

\maketitle

Artificial spin ices (ASIs) are nanostructured magnetic thin films whose magnetization textures are frustrated in an analogus way to the frustration of hydrogen bonding networks in water ice.\cite{Wang, Nisoli, Pauling}
Square ASIs are typically arrays of elongated thin film segments whose shape anisotropy forces their magnetization to align uniformly along their long axes; these therefore behave as binary Ising spins.
Despite being an array of well-defined, interacting Ising spins, an ASI does not order at low temperatures due to geometric arrangement of the array, which leads to an exponentially large manifold of ground states, none of which have net magnetization.\cite{Ramirez, Harris, YiQi, Gilbert}

Magnetic excitations of ASIs are analogous to those of natural (bulk) spin ices.\cite{Fennell1, Fennell2, Morgan, Mol}
It is predicted that one or more ``ice rule'' violating spins in ASIs create non-zero divergences in their effective magnetic fields giving rise to local effective magnetic charges.\cite{Bramwell, Castelnovo}
These magnetic charges have the qualities of quasiparticles that reflect the behavior of hypothetical magnetic monopoles.\cite{Jaubert}
Since it costs no energy to flip pairs of neighboring spins, the magnetic charges can be moved relatively easily the ASI; and pairs of charges of opposite polarity are connected by a chain of flipped spins, or ``Dirac strings'', which result from single Ising spin-flip excitations within a superdomain of ground state magnetic texture.\cite{Mol}  
This model has been generally verified at remanence in ASI,\cite{Mengotti, Ladak, Phatak} though the existence and behavior of monopoles in a square ASI at an applied magnetic field condition has not been studied in any detail.

Herein we report resonant coherent soft x-ray scattering measurements which demonstrate that partially magnetized two dimensional (2D) square ASIs form magnetic superdomains, which are in some cases bounded by magnetically charged domain walls.  
The superdomains and the charged walls can extend over tens of lattice sites, are stable under applied fields, and change shape and location based on the field history of the ASI. 
Specific superdomain and magnetic charge structures give rise to particular complex structures near the Bragg peak positions in resonant magnetic coherent x-ray scattering\cite{Hill-McMorrow}, which makes this technique a revealing tool to characterize large-scale magnetic structures in an ASI. 
Our results suggest that the dimension, topology, and field history of an ASI can be varied to tailor ordered arrays of magnetic charge defects.

\begin{figure}
\includegraphics[width= 8.6 cm]{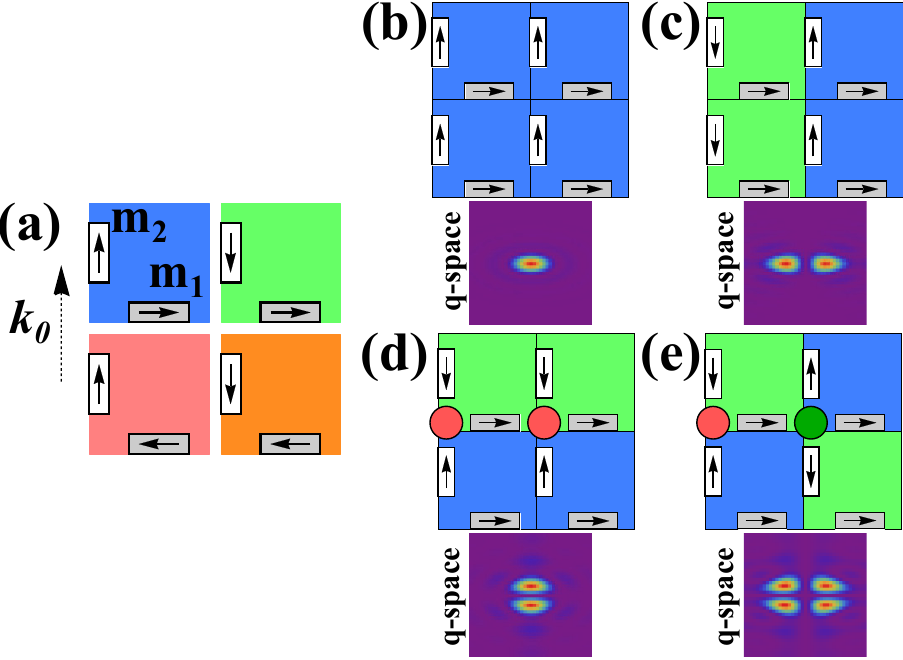}
\caption{ 
(a) Four magnetic unit cells, each containing two rectangular Ising spin islands with moments $m_1$ and $m_2$, which can be tiled to form a T2 spin ice lattice. 
Ising spins $m_i$ attain values of $\pm$1 according to their moment direction. 
Unit cell colors correspond to particular $(m_1, m_2)$ combinations. 
The x-ray scattering process is sensitive only to the component of magnetization that is collinear to the incident beam direction ($\mathbf{k_0}$). 
We have shaded nanoislands to which the x-ray scattering is not sensitive with light gray. 
(b)-(e) Different T2 superdomain textures and their corresponding scattering patterns. 
Depending on the superdomain interfaces, either single Bragg peaks (b), split Bragg peaks (c, d), or a Bragg peak split into four spots (e) can obtain.
}
\label{F1} 
\end{figure}

In a 2D square ASI the four islands that form a vertex can adopt 16 distinct local spin configurations which can be assigned to one of four energy states, called T1, T2, T3 and T4. (See Supplementary Material.\footnote{See Supplemental Material below for brief descriptions of ASI energy states T1-T4 and the sample preparation process, as well as a derivation of the magnetic structure factor.}) 
Analogous to the ``two-in/two-out'' ice rules for the atomic moments in a tetrahedral spin ice structure, T1 states, with one pair of opposing magnetic moments pointing inward and one pointing outward  from a vertex, have the lowest magnetostatic energy at zero applied field.  
T2 states also have  two-in/two-out local spin textures but with pairs of opposing inward and outward moments. 
T2 states have higher energies than T1 states because of the inequivalent distance between the four nanobars which introduces asymmetry in the interaction energy between the four elements of a vertex.
States having three (T3) or four (T4) spins pointing in or out, respectively, exhibit higher magnetostatic energies and net magnetic flux (i.e., local magnetic monopoles).  
Generally, after multiple field cycling, 2D square ASIs attain states that are primarily mixtures of T1 and T2 configurations with sporadic T3 vertices.\cite{Farhan}  

The sensitivity of resonant coherent x-ray scattering to magnetic domain morphology \cite{Seu, Chen} means that the scattered signal will significantly change depending on whether the beam encounters magnetic charges in the sample. 
Using formalism of resonant x-ray magnetic scattering we performed model scattering calculations for different T2 domain structures in 2D square ASIs. 
For a T2 spin texture, we can build magnetic structures using a set of four square ASI unit cells, each containing two rectangular Ising spin islands (see Fig. 1(a)). 
The rectangular islands are labelled by terms $m_1$ and $m_2$, which can be assigned values of $\pm$1 according to their Ising moment directions. 
Taking all possible combinations, we have four different magnetic unit cells with four different T2 bases, which we identify by four colors in Fig. 1(a).
Each kind of unit cells can be tiled to create specific T2 superdomain structures, and the complete set of square magnetic unit cells allow us to generate mixtures of all possible T2 superdomain structures. 

Figs. 1(b)-1(e) schematically show simple representative T2 domain textures, with Ising island $m_2$ along to the incident beam ($\vec{k}_0$), and their corresponding calculated scattering profiles.
In the calculation we have neglected the spins that lie transverse to the beam propagation direction (colored grey in Fig. 1) because resonant x-ray scattering is not sensitive to magnetization components transverse to $\vec{k}_0$. 

Scattering from T2 superdomains is modelled by unit cell structure factors, $S_{m_1,m_2}(\vec{q})$.\footnotemark[\value{footnote}]
T2 superdomains in the illuminated area are treated as sums over the lattice points $\vec{r}_{i,j}$ encompassed by each superdomain: 
$\sum_i \sum_j e^{i \vec{q} \cdot \vec{r}_{i, j}} S_{m_1, m_2}(\vec{q})$. 
For cases in which the transverse magnetization,  $m_1$, can be neglected, we obtain a T2 unit cell magnetic structure factor of:
\begin{align*}
S_{m_1, m_2}(\vec{q}) &\propto \text{sinc} \left( \frac{ t q_z }{ 2 } \right) 
e^{ i t q_z / 2 + i w \left( q_x + q_y \right) / 2 } \times \\
&m_2 \cos (\theta ) e^{i l q_y} \text{sinc} \left( l q_y / 2\right) \text{sinc}\left( w q_x / 2 \right),
\end{align*}
where $\theta$ is the incidence angle; $l$ and $w$ are, respectively, the width and length of an Ising nanoisland; and $q_i$ is a momentum transfer component along the ASI lattice vector $\vec{a}_i$.
The scattering intensity is proportional to
$
\lvert
\sum_{ d } (S_{m_1, m_2, d}(\vec{q}) \times \sum_{ \vec{r}_{i, j, d} }  e^{i \vec{q} \cdot \vec{r}_{i, j, d}})
\rvert^2.
$
The sum over $d$ runs over different regions of the ASI, such as superdomains; the sum over lattice points $\vec{r}_{i, j, d}$ are within region $d$.

Specific Bragg peak structures in our calculations are associated with particular superdomain structures and the presence of magnetic charge at interfaces between these superdomains.
For example, when all the spins are aligned (\textit{i.e.}, a saturation state with a global T2 domain and no magnetic charges), a well defined Bragg peak is obtained, as in Fig. 1(b).
Figs. 1(c) and 1(d) show two types of superdomain interfaces; along (Fig. 1(c)) and transverse (Fig. 1(d)) to the beam direction.
Magnetic charges, represented by red dots, form in the case of Fig. 1(d). 
Even though our calculation shows the Bragg peak forming a split structure in both cases, a split along the beam propagation direction indicates that magnetic charges at the superdomain interface.
On the other hand, a split transverse to the beam direction indicates the lack of magnetic charges. 
In addition, a four-fold intensity pattern associated with an ASI shown in Fig. 1(e), in which the corners of blue and green superdomains meet, have a mix of positive and negative magnetic charges (green and red dots).
It is important to note that the geometry shown (x-ray beam along a principal axis) permits us to detect magnetic charges but limits information about which charges are positive or negative.
(The charge assignment in Fig. 1(e) is arbitrary.)
Nevertheless, Bragg peak profiles in coherent x-ray scattering are indicators of the presence of magnetic charges and their distribution in the ASI.

Using this framework, we examined the magnetic texture of a 2D square ASI made of Permalloy (Ni$_{0.81} $Fe$_{0.19}$) nanoislands on a silicon wafer. 
The nanoislands were fabricated using electron beam lithography and had a  thickness \emph{t} = 25 nm, width \emph{w} = 50 nm, length \emph{$\ell$} = 150 nm, and with a lattice constant \emph{d} = 300 nm.
Details of sample fabrication are given in the Supplementary Material.\footnotemark[\value{footnote}]

Resonant coherent soft x-ray scattering was performed at Beamline 12.0.2.2 at the Advanced Light Source, Lawrence Berkeley National Laboratory. 
To make $\vec{k}_0$ as near to coplanar as possible with the nanoisland magnetizations, we used a reflection geometry at a low grazing incidence angle of 9$^{\circ}$. 
The coherent x-ray beam was obtained by placing a 10 $\mu$m pinhole at the monochromator focus, located 5 mm upstream from the ASI.
We used a charge-coupled device camera placed 0.5 m downstream of the ASI to record diffraction patterns at several applied magnetic fields. 
Our coherent $\sigma$ polarized incident x-ray beam was resonantly tuned to the Fe  L$_3$ edge (707 eV) to enhance the magnetic contrast. 
Since on resonance $\sigma$ polarized soft x-rays are sensitive to the component of magnetization along the beam direction (\textit{i.e.}, $\vec{k_0}$$\parallel$$\vec{M}$) \cite{Kortright}, we applied magnetic fields along the beam direction to manipulate the magnetization of the ASI.  

\begin{figure}
\includegraphics[width= 8.6 cm]{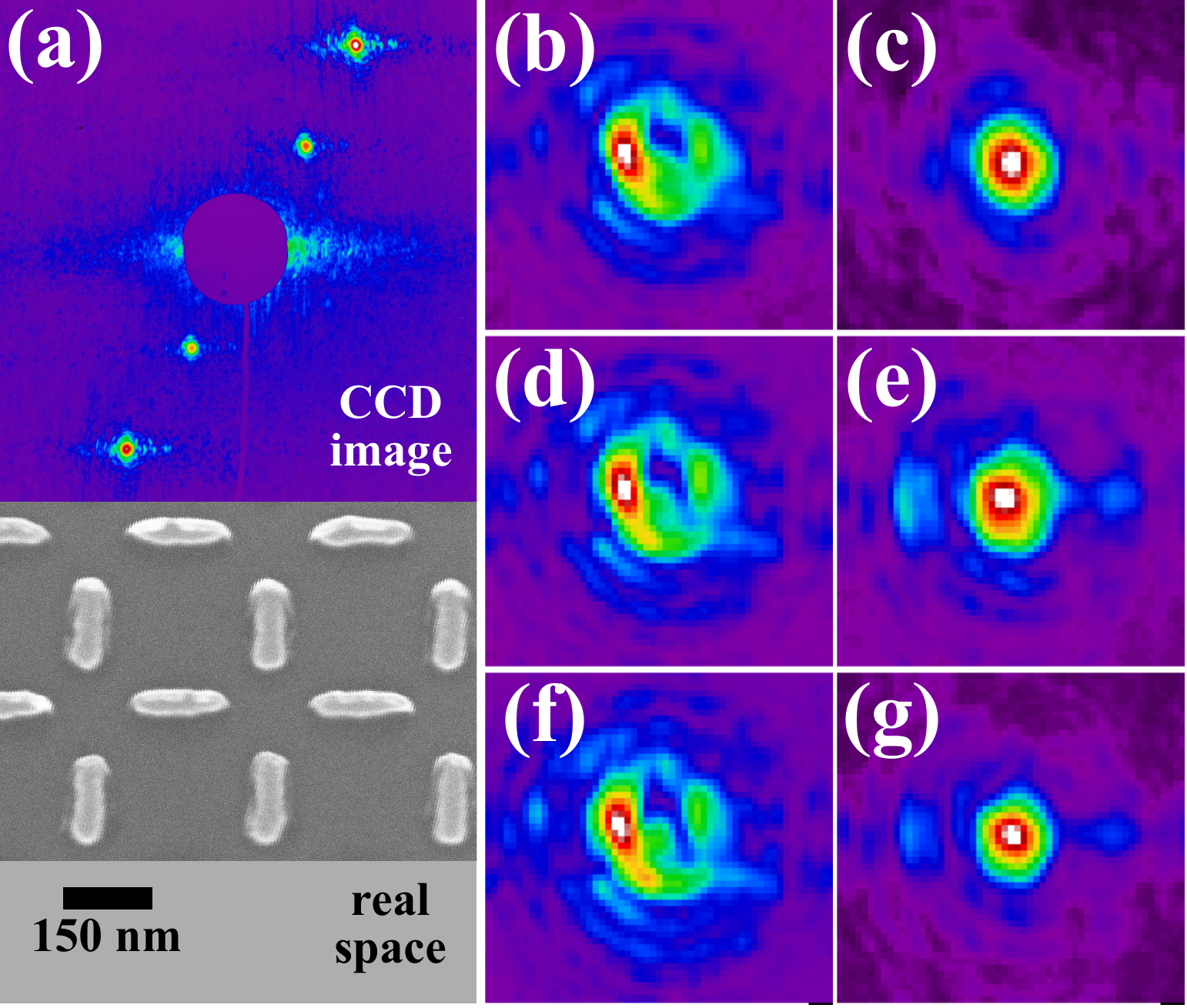}
\caption{ 
(a) Resonant coherent x-ray diffraction (top) and a scanning electron micrograph (bottom) from the square ASI.
The row of Bragg peaks is slanted because of a $\approx$ 5$^{\circ}$ misalignment between $\vec{k_0}$ and $\langle$10$\rangle$ directions. 
Parts (b), (d), and (f) show complex Bragg peak structure while magnetic fields of strengths B = 7.9 mT, B = 24.9 mT, B = 50 mT, respectively, were applied, becoming two intensity arcs.
This general shape persists till B $\approx$ 80 mT, though the arcs subtly change as the field is increased.
Parts (c), (e), and (g) show approximately Gaussian Bragg peaks after the magnetic field is turned off.    
}
\label{F2}
\end{figure}

Fig. \ref{F2}(a) shows a series of Bragg peaks recorded in zero field conditions from the 2D ASI.  
Well formed, nearly Gaussian Bragg peaks can be seen, accompanied by diffuse Airy fringes caused by diffraction from the circular pinhole.
The Gaussian Bragg peaks indicate that there is no contrasting pattern to the nanoisland magnetic moments that could cause fine structure to form in the Bragg peaks in zero field.
Moreover, the diffraction pattern indicates that the magnetic texture has the same period as the lattice of Permalloy nanoislands.

Figs. \ref{F2}(b)-\ref{F2}(g) illustrate what happens when magnetic fields are applied to the ASI.
Before recording the applied magnetic field dependent data, we saturated the sample in the direction opposite to the beam propagation direction and then turned off the field. 
The intensity pattern is remarkably different when a magnetic field is applied (Figs. \ref{F2}(b), 2(d), 2(f)). 
Instead of single Gaussian peaks, we observe areas of low intensity surrounded by a higher intensity ring forming at the Bragg peak positions. 
The rings are not uniformly circular: two strong intensity regions form along the scattering plane, connected by weaker arcs of intensity.
We also note that all the Bragg peaks have similar structures, the sizes of which do not increase with diffraction order.  
This Bragg peak structure is observed over the measured field range of 10 mT $ \leq \mu_0$H$_z$ $\leq$ 80 mT.
As discussed above and shown in Fig. \ref{F1}, the Bragg peak splitting along the beam direction indicates that the x-rays interacted with magnetically charged superdomain walls. 
Further, on switching off the magnetic field, we again observe Gaussian Bragg peaks (Figs. 2(c), (e), (g)) which means that the magnetically charged walls are stable when the field is applied but the spin texture changes as soon as the field is removed. 

Creation of a new \emph{long range ordered} magnetic structure with a different period than the nanoisland lattice can be excluded since that would produce new peaks either between existing Bragg peaks or distinct satellites around the Bragg peaks, rather than the observed scattering pattern.  
This therefore suggests that there is arrangement of charged interfaces between superdomains that causes Bragg peak to form a specific fine structure.

\begin{figure}
\includegraphics[width= 8.6 cm]{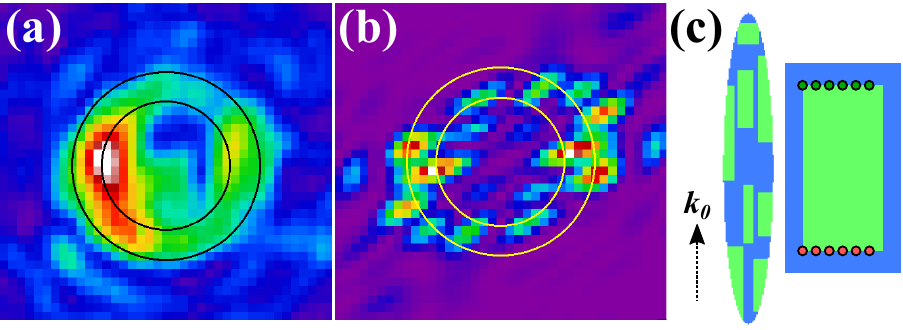}
\caption{ 
(a) Detailed view of a measured Bragg peak. (b) Calculated Bragg peak based on the real space model shown in (c), the real space model with T2 superdomains.  Considering the spin texture as shown in Fig. 1(a), magnetic charges of opposite polarity form along the superdomain interface transverse to beam direction indicated by $\vec{k}_0$.
}
\label{F3}
\end{figure}

While the zero field ground state of square ASIs is one with solely T1 vertices, macroscopic T1 states rarely form due to thermalization issues.  
Coexistence of superdomains made of T1 and T2 states is consistent with the presence of large energy barriers ($\approx$ 10$^5$ K) that suppress full equilibration of the micrometer-scale ASI film segments at room temperature\cite{Nisoli,Heyderman}, which is far below the ferromagnetic phase transition of a 25 nm-thick permalloy film.\cite{Mauri}
Moreover, a moderate in-plane applied magnetic field breaks the ideal zero-moment T1 symmetry and favors T2 states with net moments aligned along the field, as observed in PEEM.\cite{Farhan}
The magnetic field history will therefore largely determine the degree of order and the energy of the magnetic texture attained by the square ASI samples studied here.
During our experiments, we drove the ASI through hysteresis loop between maximum fields of $\pm 0.4$ T that likely introduced a heterogeneous texture composed of T1 and T2 states.
We expect that a square ASI with partial in-plane magnetization to have an increasing fraction of T2 states after saturating the ASI with greater applied fields.
This tendency will play a key role in the degree of local magnetic ordering and magnetic superdomain formation.

To ascertain which T2 structures are consistent with the observed diffraction pattern, we created model scattering structure factors corresponding to partially magnetized square ASI lattice.
Candidate magnetic structures were built of unit cells like those shown in Fig. 1(a).
The unit cell and two rectangular Ising nanoislands within it have the same nominal dimensions as those of the sample ASI. 
We considered magnetic structures that reside in the illuminated area of the sample.
Due to the 9$^{\circ}$ incidence angle, the illuminated area resembles an ellipse with its major axis parallel to the beam propagation direction that illuminates $\approx$ 600 ASI unit cells (Figs. 2(b) and 2(c)). 
We note that due to non-sensitivity towards transverse spins, the green (red) unit cell shown in Fig. 1(a) is indistinguishable from orange (blue) unit cell.  
Thus, it is sufficient to consider only the case with blue/orange or blue/green unit cell mixtures, as these two combinations yield identical scattering patterns.


The superdomain structures shown in the ellipse in Fig. 3(c) produce the calculated scattering pattern in Fig. 3(b)). 
Although simple, our calculated intensity profile captures the important features of the data (Fig. 3(a)). 
We have reproduced the intensity arcs as well as the two higher intensity lobes. 
In addition the calculated intensity pattern follows the same pixel numbers as in the data (the ring in Fig. 3(a) and (b) show equal pixel contours), which points to a statistically correct distribution of superdomains with magnetically charged interfaces.

Detailed views of the superdomain interface is shown to the right side of the ellipse in Fig. 3(c).
Recalling the spin configuration for the green and blue unit cell (Fig. 1(a)), the superdomain interface transverse to $\vec{k}_0$ has magnetic charges. 
No magnetic charges are present in the superdomain wall along the beam direction. 


The formation of superdomain walls in a Permalloy thin film and the homogeneity of the magnetization near the sharper ends of the ASI segments are affected by the size of the film thickness.  
In addition, the T2 superdomain morphology will vary from sample to sample, and perhaps from position to position on a single sample.  
We therefore investigated if emergent magnetic charges appear in other square ASI of different thickness, or are unique to the sample that we studied.
We repeated the scattering experiment with a square ASI with film segments of 3 nm thickness, and the resulting scattering pattern, which is remarkably different than what we observed for 25 nm thick film, is shown in Fig. 4(a).
\begin{figure}
\includegraphics[width= 8.6 cm]{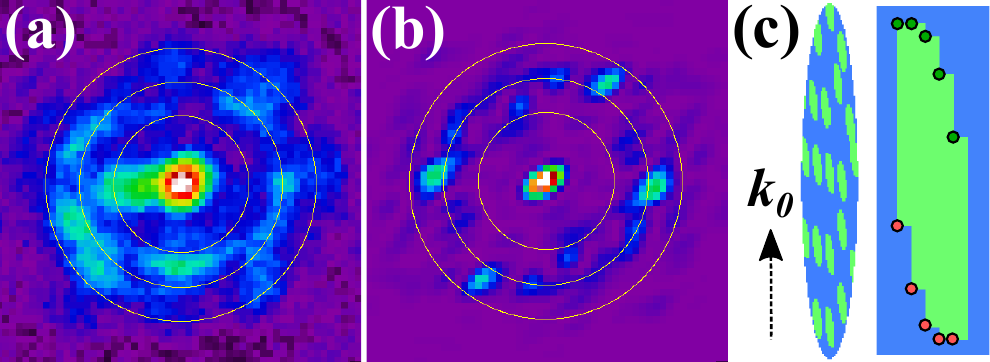}
\caption{ 
(a) Coherent x-ray scattering profile from an as grown square ASI with 3 nm thick Ising islands. 
A sharp central peak is surrounded by a ring of spots. 
(b) Calculated coherent x-ray scattering pattern: a central peak surrounded by a ring of intensity spots. 
The circular lines in (a) and (b) are equal-pixel distance contours. 
Calculated intensity patterns agree well with the observed pattern.  
(c) Real space model corresponding to the scattering pattern in (b).
}
\label{F4}
\end{figure}

The Bragg peak profile is a central bright spot that is surrounded by a ring of intensity spots.
The scattering pattern in Fig. 4(b) that closely resembles the experimental data arises from the T2 superdomain structure in Fig. 4(c).
The superdomains in the thinner ASI are smaller in size and are dispersed through out the lattice.
Interfaces with head-to-head (or tail-to-tail) Ising spin textures will give rise to local divergences in the magnetization and corresponding magnetic charges.


Extended superdomain wall structures containing emergent magnetic charges could plausibly be used to create magnetic charge circuits,\cite{ Greedan, Vedmedenko} using the superdomain walls as magnetic charge paths and the ASI geometry tuned to determine the circuit topology.
ASIs with thin nanoislands, in which we have seen signs of spontaneous charge motion, may be amenable to this. 
Among other properties, the characteristics of magnetic charge motion in ASIs can be measured by studying the time evolution of Bragg peak structure.

\footnotetext{See Supplemental Material at [URL will be inserted by publisher] for brief descriptions of ASI energy states T1-T4 and the sample preparation process, as well as a derivation of the magnetic structure factor.}

\begin{acknowledgments}
Work at ALS, LBNL was supported by the Director, Office of Science, Office of Basic Energy Sciences, of the US Department of Energy (Contract No. DE-AC02-05CH11231). J.C.TL., R.S., P.F. and S.D.K. acknowledges support by the Director, Office of Science, Office of Basic Energy Sciences, Materials Sciences and Engineering Division, of the U.S. Department of Energy under Contract No. DE-AC02-05-CH11231 within the Nonequilibrium Magnetic Materials Program  (KC2204).
Work at ANL is supported by the U.S. Department of Energy, Office of Science, Office of Basic Energy Sciences, under Contract DE-AC02-06CH11357. Work at the Unversity of Oregon was partially supported by the U.S. Department of Energy, Office of Basic Energy Sciences, Division of Materials Science and Engineering under Grant No. DE- FG02-11ER46831. Research at the University of Kentucky was supported by U.S. Department of Energy Grant No. DE-FG02-97ER45653 and U.S. NSF Grant No. DMR-1506979.
\end{acknowledgments}

\bibliography{manuscript.bib}

\appendix
\renewcommand\thefigure{S\arabic{figure}}
\section{Supplemental Material}
\setcounter{figure}{0}

\renewcommand{\thesection}{S\arabic{section}}
\section{Magnetic vertices in square artificial spin ices}

In a 2D square ASI the four islands that form a vertex can adopt 16 distinct local spin configurations which can be assigned to one of four energy states, called T1, T2, T3 and T4.
Analogous to the ``two-in/two-out'' ice rules for the atomic moments in a tetrahedral spin ice structure, T1 states, with one pair of opposing magnetic moments pointing inward and one pointing outward  from a vertex, have the lowest magnetostatic energy at zero applied field.
T2 states also have  two-in/two-out local spin textures but with pairs of opposing inward and outward moments. 
T2 states have higher energies than T1 states because of the inequivalent distance between the four nanobars which introduces asymmetry in the interaction energy between the four elements of a vertex.
States having three (T3) or four (T4) spins pointing in or out, respectively, exhibit higher magnetostatic energies and net magnetic flux (i.e., local magnetic monopoles).  
Generally, after multiple field cycling, 2D square ASIs attain states that are primarily mixtures of T1 and T2 configurations with sporadic T3 vertices. 
\begin{figure}[h]
\begin{center}
\includegraphics[width=8.6 cm]{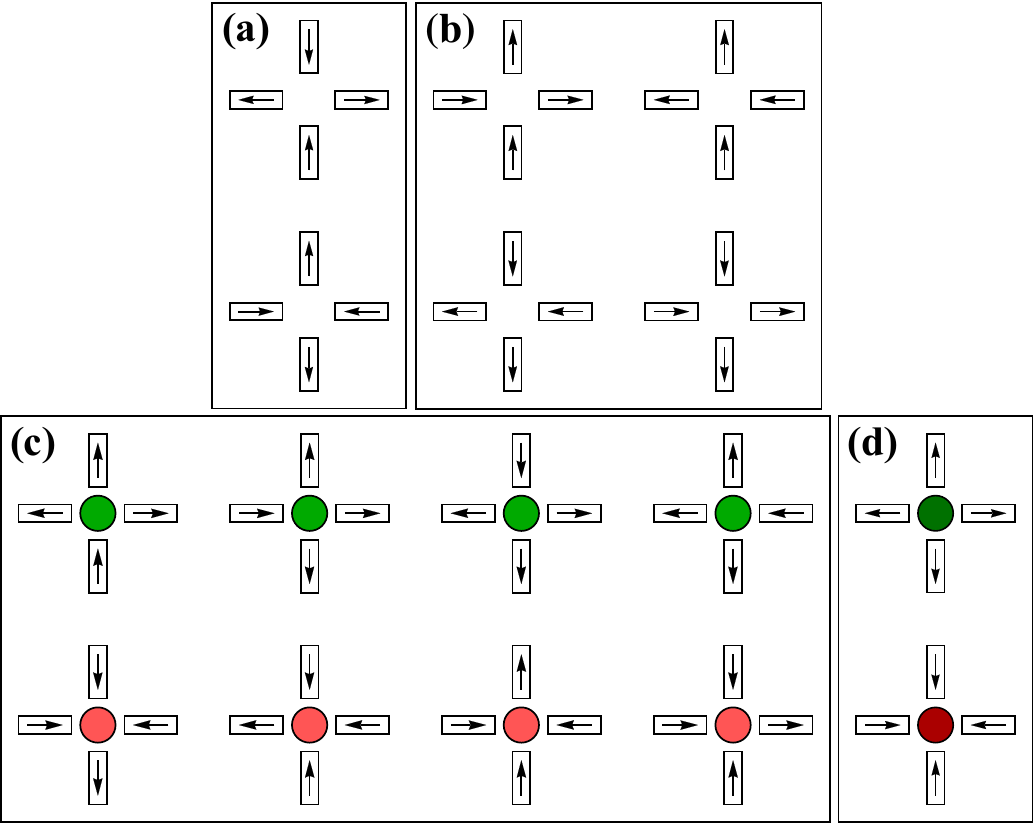}
\caption {
The Ising spin textures associated with (a) T1, (b) T2, (c) T3, and (d) T4 energy states of square ASIs. 
The two lowest energy configurations, T1 and T2, have two of four magnetic moments oriented toward the vertex and two oriented away from the vertex, analogous to the ``two in/two out'' ice rules for tetrahedral water ice. 
The two highest energy configurations, T3 and T4, display magnetic charges at their vertices, shown by green and red dots. 
}
\end{center}
\end{figure}

\section{Sample growth procedure}
\label{growth}

We have used electron beam lithography to pattern square lattices of Permalloy dots of thickness \emph{t} = 25 nm, width \emph{w} = 50 nm, length \textit{l} = 150 nm, and lattice constant \textit{a} = 300 nm. 
ZEP positive resist was spin-coated on a Si wafer prior to electron beam exposure. 
After the e-beam exposure and development, a Permalloy film of thickness 25 nm was then deposited using electron beam evaporation, with a base pressure of 10$^{-7}$ Torr. 
Final lift-off of resist was done using N-Methyl-2-pyrrolidone (NMP). 
Our sample had a 2 x 2 mm overall dimension. 
\begin{figure}[h]
\begin{center}
\includegraphics[width=8.6 cm]{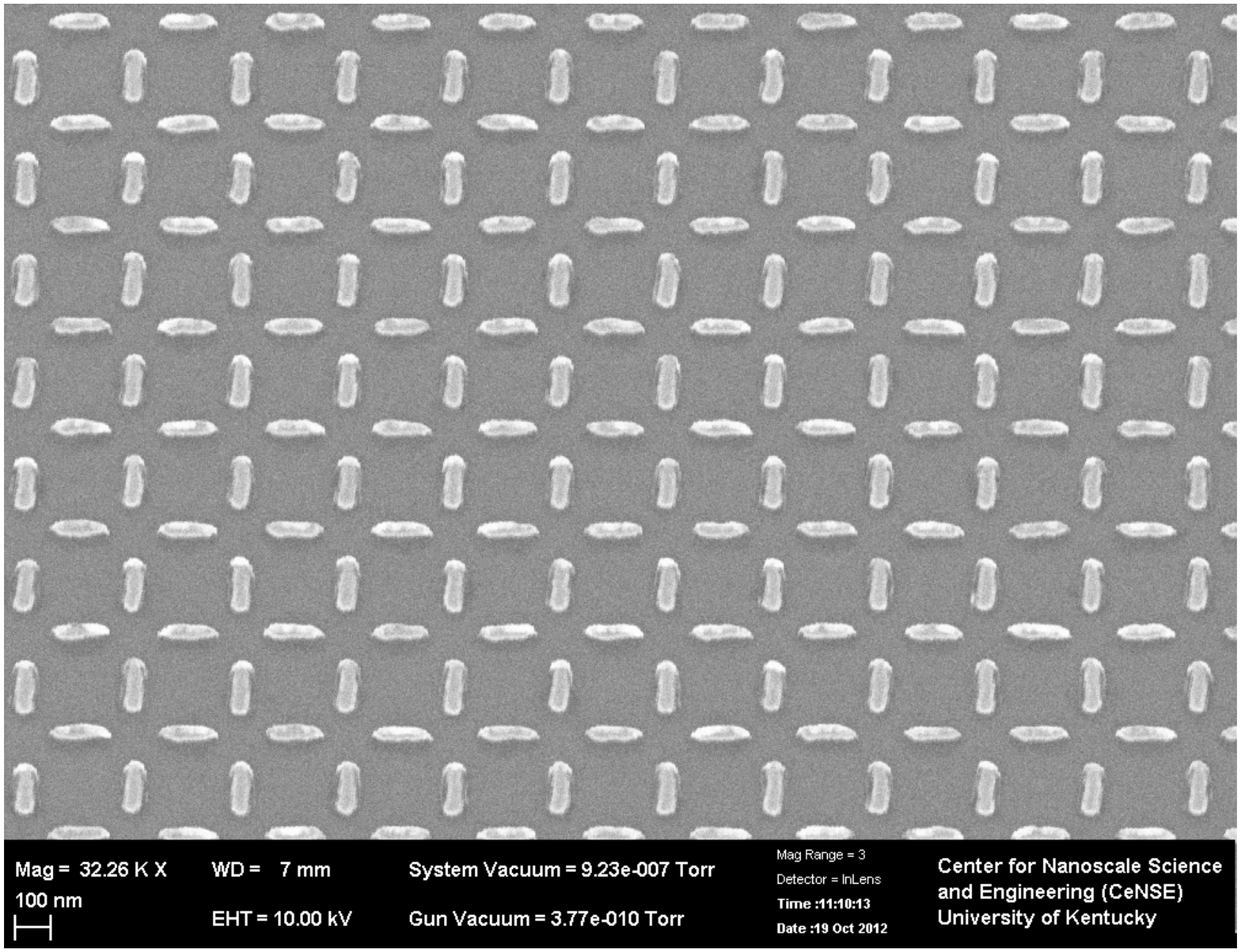}
\caption {Scanning electron micrograph of a square array of Permalloy dots of thickness \emph{t} = 25 nm, width \emph{w} = 50 nm, length \emph{$\ell$} = 150 nm, lattice constant \emph{a} = 300 nm, and total dimension of the array is 2 mm x 2 mm.}
\end{center}
\end{figure}

\section{Calculation of magnetic structure factor}

The sensitivity of resonant coherent x-ray scattering to magnetic domain morphology means that the scattered signal will significantly change when the beam illuminates magnetic charges in the ASI. 
The magnetic structure factor, being proportional to the scattered field, is necessary to understand how the resonant scattering from superdomains of the ASI will interfere with each other. 
This interference results in a far field scattered intensity that, as shown in Fig. 1 of the main text, is dependent on the magnetic state of the ASI. 
In this section, we will explain how we arrived at the structure factor used in the main text (Equation 1).

As discussed in the text, we assume that a moderate in-plane applied magnetic field breaks the ideal zero-moment T1 symmetry and favors T2 states with net moments aligned along the field.
Based on this reasoning, the square ASI will be dominated by T2 superdomains and the relevant magnetic structure factors are those of T2 states.

\begin{figure}
\begin{center}
\includegraphics[width=8.6 cm]{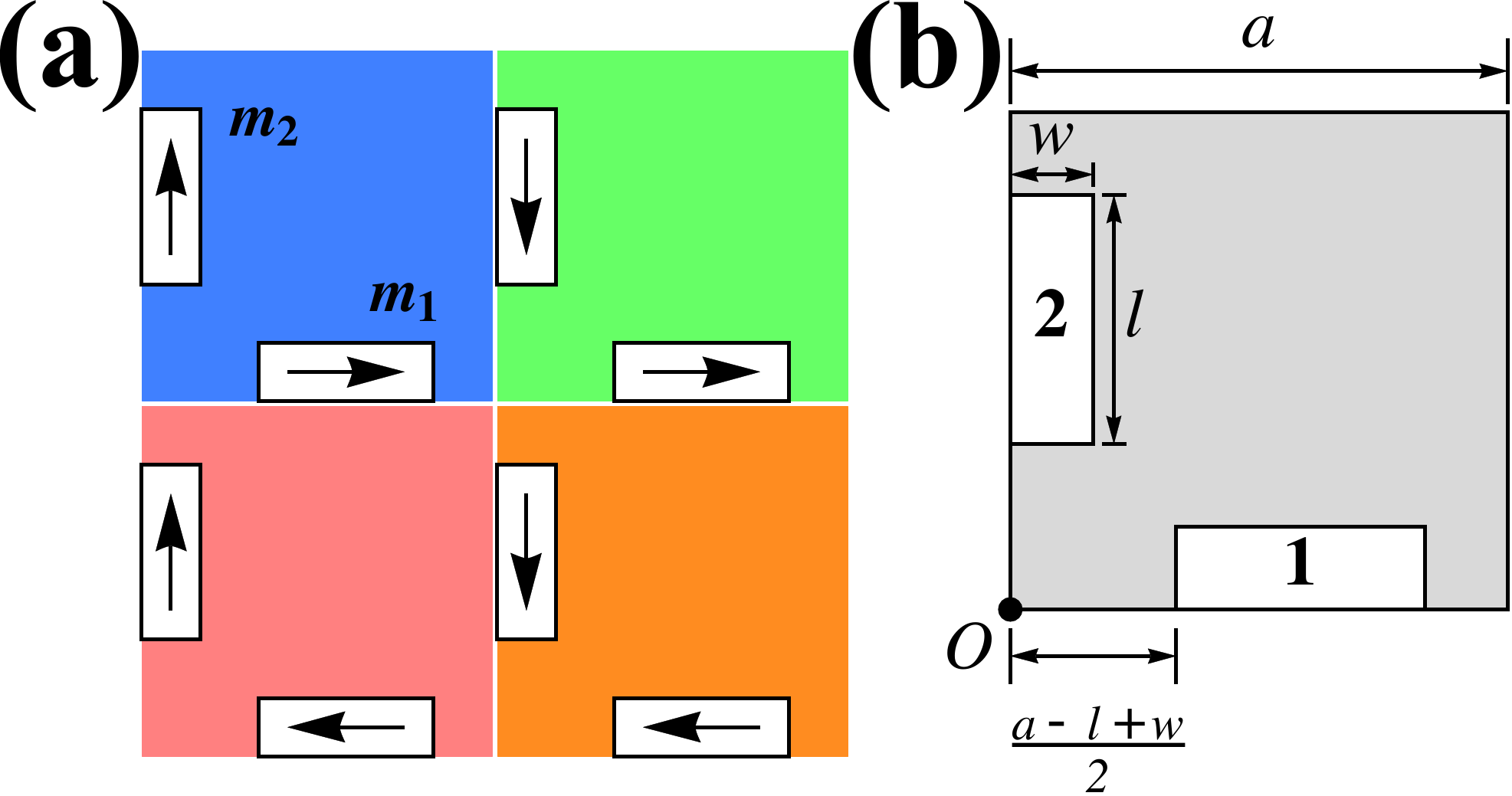}
\caption {Real space diagrams of the unit cell types of square ASIs. (a) Unit cells containing each of the four possible magnetic bases for T2 square ASI (reproduced from Fig. 1(a) in main text) and (b) the charge unit cell. Both the charge and magnetic unit cells are squares with sides as long as the lattice constant, \textit{a}. This unit cell contains orthogonally oriented Ising nanoislands with dimensions \textit{l} and \textit{w}. The nanoislands are treated as sharp edged rectangular prisms in the structure factor calculation.}
\end{center}
\label{SFfigure}
\end{figure}

Four Ising bases, shown in four different colored square unit cells in Fig. S3(a), span all possible T2 states. 
The magnetic form factors of the nanoislands will differ in the resonant magnetic scattering process due to their different magnetization directions.
Thus, there are four distinct magnetic form factors, one for each of the T2 Ising bases.
Our calculation reflects this and the general expression of the structure factor explicit depend on both $m_1$ and $m_2$.
Ultimately, due to the scattering geometry, only the $m_2$ dependent term will matter.

A lattice of like T2 unit cells shares the same translational symmetry and lattice constant as the charge lattice of the square ASI, which makes the charge and magnetic unit cells and lattice vectors identical.
However, the magnetic form factors of the nanoislands will differ in the resonant magnetic scattering process due to their different magnetization directions.

In setting up the calculation, we treated the nanoislands as rectangular prisms of uniform density with the same nominal dimensions of nanoislands in the ASI, as displayed in Fig. S3(b).
Their magnetizations are labelled by terms $m_1$ and $m_2$ which can be assigned values of $\pm$1 according to their Ising moment directions (positive if $m_1$ and $m_2$ were, respectively, along either the x or y axes, and negative otherwise).
Multiplying the nanoisland density function with the magnetic vector $\vec{m}_i$ gives the nanoisland an Ising magnetization.

Calculating the magnetic structure factor, $S_{m_1, m_2}(\vec{q})$, amounts to calculating the $\vec{q}$ Fourier component of the unit cell charge density, $\rho_i(\vec{x})$:
\renewcommand{\theequation}{S\arabic{equation}}
\begin{align}
S_{m_1, m_2}(\vec{q}) &\propto \left ( \frac{ 2 \pi }{ a } \right )^2 \frac{ 2 \pi }{ t } \times \nonumber \\
& \int_{\text{cell}} e^{i \vec{q} \cdot \vec{x} } \hat{k}_0 \cdot \left( m_1 \hat{x}  \rho_1(\vec{x}) + m_2 \hat{y} \rho_2(\vec{x}) \right) d\vec{x}.
\label{sofq}
\end{align}
The vector $\hat{k}_0$ is the incident beam direction and the projection along it is a consequence of the resonant magnetic scattering process.

Treating the nanoislands as having uniform and identical charge densites, $\rho_0$, defined by a product of unit step functions, $\Theta(x)$, the density functions for the nanoislands are: 
\renewcommand{\theequation}{S\arabic{equation}}
\begin{align}
\rho_1(\vec{x}) = \rho_0 &\Theta( x - \frac{ l +w }{ 2 } ) \Theta( \frac{ 3l +w }{ 2 } - x ) \times \nonumber \\
		        			&\Theta( y ) \Theta( w - y  ) \Theta( z ) \Theta( t - z ), \nonumber \\
\rho_2(\vec{x}) = \rho_0 &\Theta( x ) \Theta( w - x ) \times \nonumber \\ 
				 	&\Theta( y - \frac{ l +w }{ 2 } ) \Theta( \frac{ 3l +w }{ 2 } - y ) \Theta( z ) \Theta( t - z )
\label{density-functions}
\end{align}

A lattice of like T2 unit cells shares the same translational symmetry and lattice constant as the nonmagnetic charge lattice of the square ASI, which makes the charge and magnetic unit cells identical.
With this in mind and using the density functions of Eq. \ref{density-functions}, Eq. \ref{sofq} reads:
\begin{align*}
S_{m_1, m_2}(\vec{q}) &\propto 
\rho_0 \left ( \frac{ 2 \pi }{ a } \right )^2 \frac{ 2 \pi }{ t } \hat{k}_0 \cdot \bigg( \\
& m_1 \hat{x} \int_{ \frac{ l + w }{ 2 } }^{ \frac{ 3l + w }{ 2 } } e^{ i q_x x } dx \int_{0}^{ w } e^{ i q_y y } dy + \\
& m_2 \hat{y} \int_{ 0 }^{ w } e^{ i q_x x } dx \int_{ \frac{ l + w }{ 2 } }^{ \frac{ 3l + w }{ 2 } } e^{ i q_y y } dy \\
&\bigg)  \int_{0}^{ t } e^{ i q_z z} dz.
\end{align*}
Evaluating this integral (noting that we set $\rho_0 = 1$) we obtain:
\begin{align*}
S_{m_1, m_2}(\vec{q}) &\propto \text{sinc} \left( \frac{ t q_z }{ 2 } \right) e^{ i t q_z / 2 + i w \left( q_x + q_y \right) / 2 } \\
	  			&\big( m_2 \cos (\theta ) \cos (\phi ) e^{i l q_y} \text{sinc} \left( l q_y / 2\right) \text{sinc}\left( w q_x / 2 \right) \\
				&-m_1 \cos (\theta ) \sin (\phi ) e^{i l q_x} \text{sinc}\left( l q_x / 2 \right) \text{sinc}\left( w q_y / 2 \right) \big),
\end{align*}
in which $\vec{q} = 2 \pi ( \frac{ H }{ a } \hat{x} + \frac{ K }{ a } \hat{y} + \frac{ L }{ t } \hat{z} )$ and $\hat{k}_0 = \hat{ x } \sin( \phi ) \cos( \theta ) + \hat{ y } \cos( \phi ) \cos( \theta ) + \hat{ z } \sin( \theta )$. 
The angle $\theta$ is the incident angle and $\phi$ takes into account any misalignment between the y-axis of the ASI and the scattering plane.
Since $\phi \approx -5^{\circ}$ in the present case, the $m_1$ term is negligible and $\cos( \phi ) \approx 1$, leading to Eq. 1 in the text:
\begin{align}
S_{m_1, m_2}(\vec{q}) &\propto \text{sinc} \left( \frac{ t q_z }{ 2 } \right)
e^{ i t q_z / 2 + i w \left( q_x + q_y \right) / 2 } \times \nonumber \\ 
&m_2 \cos (\theta ) e^{i l q_y} \text{sinc} \left( l q_y / 2\right) \text{sinc}\left( w q_x / 2 \right).
\end{align}

\end{document}